\documentclass[reprint,amsmath,amssymb,aps,groupedaddress,floatfix]{revtex4-1}
\usepackage{graphicx}
\usepackage{dcolumn}
\usepackage{bm}
\usepackage{natbib}
\usepackage{color,soul}
\usepackage{hyperref}

\begin{document}
\preprint{APS/123-QED}
\title{Significant reduction of lattice thermal conductivity in suspended graphene by charge doping}

\author{Ajit Jena}
 \altaffiliation[Also at ]{Key Laboratory of Optoelectronic Devices and Systems of Ministry of Education and Guangdong Province, College of Optoelectronic
Engineering, Shenzhen University, Shenzhen 518060, China}
\author{Wu Li}%
\affiliation{%
 Institute for Advanced Study, Shenzhen University, Shenzhen 518060, China \
}%
\date{\today}

\begin{abstract}
Declining the lattice thermal conductivity in graphene is essential for its thermoelectric applications. In high electron density systems, scatterings of phonons by electrons are no less than the phonon scatterings by other phonons. With the aid of first-principle calculations we examine the lattice thermal conductivity in graphene by inducing electron-phonon scattering externally. With experimentally tunable charge carrier density we find $\sim 50\%$ reduction of the lattice thermal conductivity at 200 K. The present study opens up new avenues for potential thermoelectric applications of graphene.
\begin{description}
\item[Keywords]
graphene, charge doping, electron-phonon coupling, p-p scattering, p-e scattering, lattice thermal conductivity
\end{description}
\end{abstract}

\maketitle


\section{\label{sec:level1}introduction}
The extremely large lattice thermal conductivity of suspended graphene (2600-5300 W/mK) \cite{gra-thermal1, gra-thermal2, gra-thermal3} at room temperature makes the material unarguably a suitable candidate for thermoelectric power generator. Its electronic thermal conductivity is nearly $1/10^{th}$ of the lattice contribution \cite{gra-elec-thermal} but still comparable to that of the typical metals. Being a two-dimensional (2D) system it offers physical properties that can be considerably tuned by external gate voltage. One such example is Bloch-Gr\"uneisen temperature, $\Theta_{BG}$. Efetov and Kim have shown that $\Theta_{BG}$ in graphene can be tuned up to $\sim$ 1000 K with high carrier density ($n = 4 \times 10^{14}$ cm$^{-2}$) by applying gate voltage \cite{gra-expt}. However, the external carrier density has negligible effect on the resistivity and electronic thermal conductivity \cite{gra-expt, gra-res, gra-elec-thermal}. The fundamental and practical usefulness of electron or hole doping in graphene has received much attention and therefore the subject has offered studies that are quite big in number. The novel work are mostly on tuning the electron-phonon coupling strength and how it drives the electronic transport in graphene \cite{gra-expt, eph1, eph2, eph3}.   

At the same time it deems important to examine how the lattice thermal conductivity responds to the external means, doping and defects. As it is mentioned earlier the high thermal conductivity of graphene fulfills the requirement of an efficient thermoelectric power generator. However, its application in thermoelectric is limited by large thermal conductivity and low Seebeck coefficient. One way to enhance the Seebeck coefficient and declining thermal conductivity is to modify the band structure at Fermi energy. The present study is mainly focused on the objective of reducing the lattice thermal conductivity of pristine graphene. Disproportionately large drop in lattice thermal conductivity is shown in graphene with isotopic defects \cite{latt-kappa1}. Even a small concentration of silicon impurity significantly reduces the thermal conductivity of graphene \cite{latt-kappa2}. Though there are reports on the effect of doping by foreign elements and defects \cite{latt-kappa1, latt-kappa2, latt-kappa3} on the lattice thermal conductivity the impact of charge doping is still missing in the literature. Applying external voltage is an alternative method of changing the charge carrier density effectively in low-dimensional systems. The method has been successful, particularly in graphene, in revealing the uniqueness of Dirac cone linked to unconventional integer quantum Hall effect \cite{novoselov2, kim}.

The lattice thermal conductivity ($\kappa_{ph}$) is dominated by the phonons and is defined as $\kappa_{ph} = (1/3) c_{ph}v_{ph}l_{ph}$, where $c_{ph}$, $v_{ph}$ and $l_{ph}$ are respectively the specific heat, group velocity and mean free path of the phonons. In the low temperature limit, below the Debye temperature ($\Theta_{D}$), the phonons are scattered by the impurities making $l_{ph}$ temperature independent and $\kappa_{ph}$ varies as $c_{ph}$ ($c_{ph} \propto T^2$ for 2D and $c_{ph} \propto T^3$ for bulk systems). When $T > \Theta_{D}$, $c_{ph}$ obeys Dulong-Petit law where phonon-phonon (p-p) anharmonic scattering is dominated and $l_{ph}$ varies as $1/T$ implying $\kappa_{ph} \propto 1/T$. We calculate the $\kappa_{ph}$ of graphene by solving the phonon Boltzmann transport equation (BTE) using p-p anharmonic inter atomic force constants and phonon-electron (p-e) scattering rates. P-e scattering rates are included to incorporate the effect of charge doping on lattice thermal conductivity. 
We show that with doping the phonon life time is reduced leading to the fall in thermal conductivity. We find that with large electron doping, that is achievable in experiment, the lattice thermal conductivity in graphene can be reduced by $\sim 50\%$ at 200 K.  

\section{\label{sec:level2}computational methodology}
Boltzmann Transport Equation has been proven successful in describing the heat transport phenomena in materials \cite{lindsay1, lindsay2, shengbte}. We use modified-ShengBTE package \cite{shengbte} to calculate $\kappa_{ph}$ by supplying second and third order inter atomic force constants along with p-e scattering rates obtained from the electron-phonon Wannier (EPW) calculations \cite{epw}. For EPW, we solve BTE accurately \cite{wuli} using Allen's model \cite{allen}. We employ pseudo-potential based density-functional theory (DFT) and density-functional perturbation theory (DFPT) as implemented in Quantum ESPRESSO \cite{giannozzi} within the framework of local density approximation (LDA) to compute the electron energies, force constants and e-ph matrix elements. The matrix elements are calculated first on a coarse grid of $8\times8\times1$ and then Wannier interpolated into a fine gird of $200\times200\times1$. 
Norm-conserving pseudo-potential is used in the calculations and the kinetic energy cutoff for the planewave is taken as 60 Ry. The electronic integration over the Brillouin zone is approximated by the Gaussian smearing of 0.025 Ry for the self-consistent calculations. The third order force constant calculation is performed on supercells containing $8\times8\times1$ unit cells including up to ten nearest-neighbors interaction. Finally, all the necessary inputs are provided to ShengBTE code to calculate $\kappa_{ph}$ on a q-grid of $140\times140\times1$. The graphene sheets are sufficiently isolated from each other by 10 \AA{} of vacuum to ensure the negligible interlayer interaction.

\section{\label{sec:level3}results and discussions}

\begin{figure}
\centering
\includegraphics[scale = 0.32]{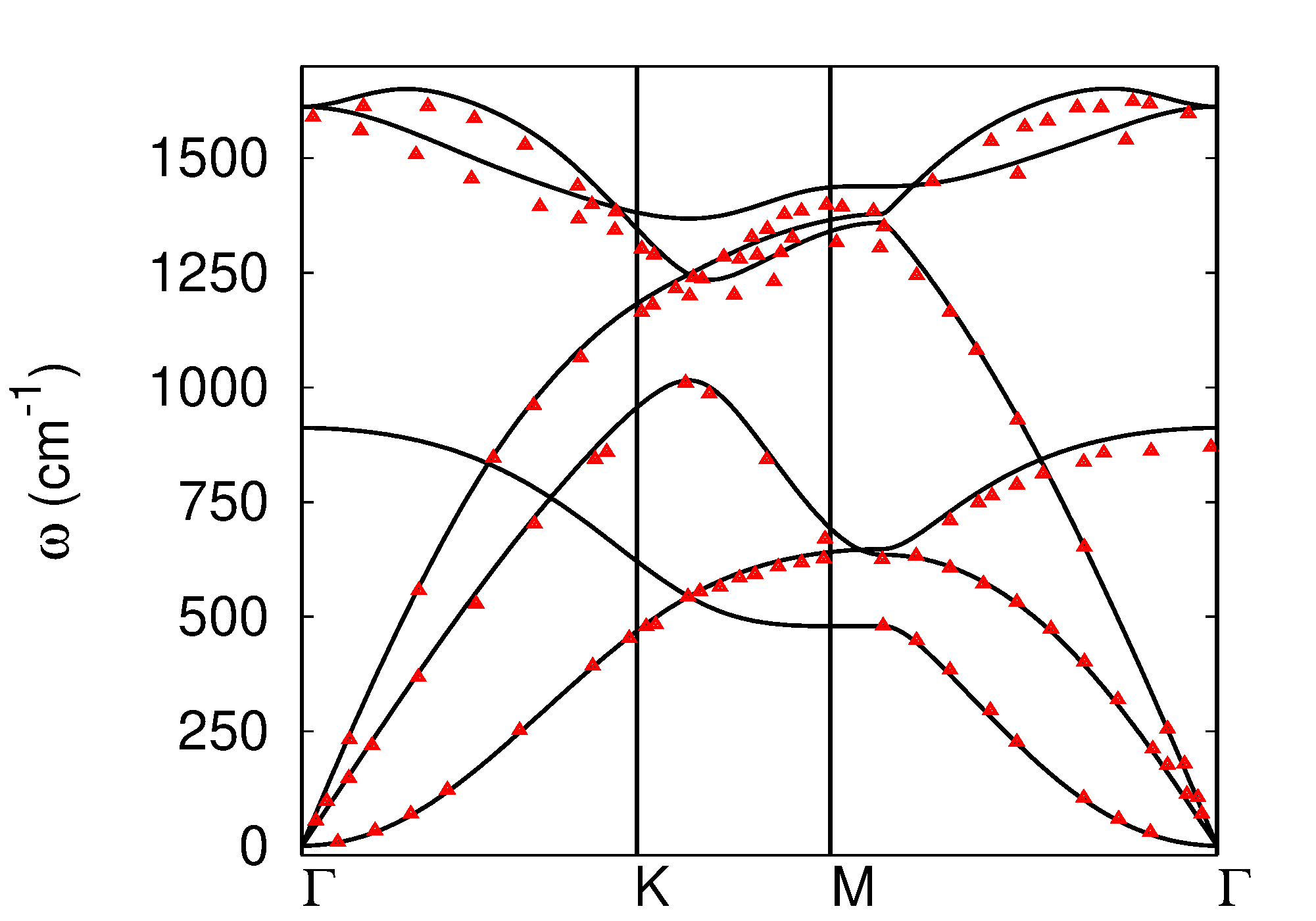}
\caption{Phonon dispersion relation of graphene along the high symmetry-lines of hexagonal lattice. The triangles represent the experimental data from \cite{mohr} obtained by inelastic x-ray scattering for in-plane graphite.}
\label{fig:phdis}
\end{figure}

\begin{figure}
\centering
\includegraphics[scale = 0.32]{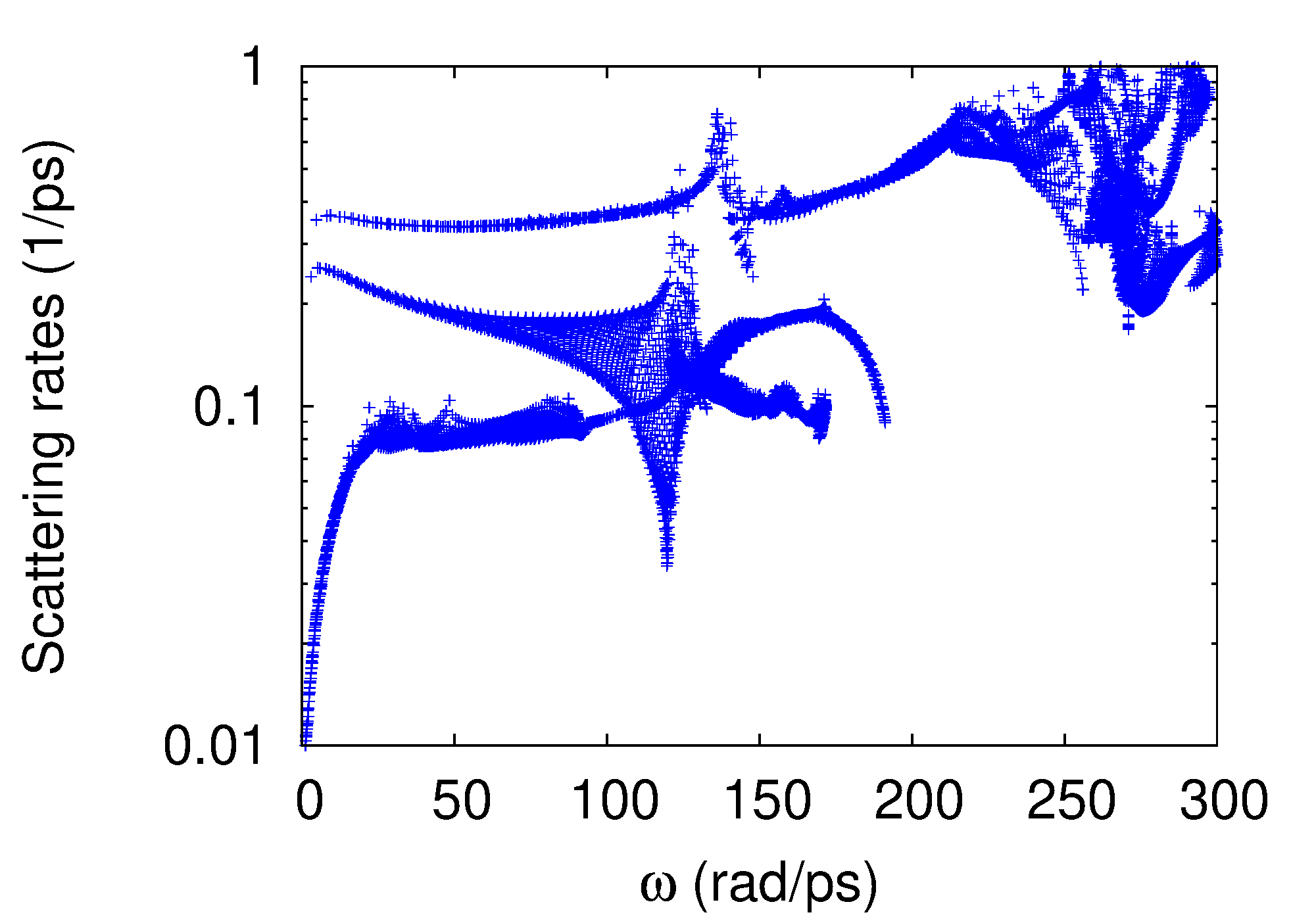}
\caption{Room temperature phonon-phonon anharmonic scattering rates of pristine graphene.}
\label{fig:gra-pp}
\end{figure}

Since phonons are the sole carriers for lattice heat conduction it is important to calculate the phonon frequencies accurately. In Fig.~\ref{fig:phdis}, we have presented the phonon dispersion of graphene along the high symmetery lines of hexagonal lattice. In the long wavelength limit, the in-plane transverse acoustic (TA) and longitudinal acoustic (LA) modes obey the normal linear dispersion and the out-of-plane acoustic (ZA) mode shows $q^{2}$ frequency dispersion, typical in 2D materials, which is a consequence of $D_{6h}$ point group \cite{d6h}. The experimental phonon frequencies obtained from inelastic x-ray scattering for in-plane graphite are also compared in the Figure. Fig.~\ref{fig:phdis} demonstrates that our first principle based DFPT harmonic inter atomic force constant calculations have good agreement with the experimental results. Further to validate our results, the p-p anharmonic scattering rates are plotted in Fig.~\ref{fig:gra-pp}. For long wavelength, while the ZA scattering rates are proportional to $q^{2}$ the scattering rates of LA and TA modes approach constant values as reported in references \cite{bonini, wuli-gra}. Hence, it is expected that our calculations will produce the properties, discussed in this study, accurately that are yet to be realized in experiment.

In addition to the p-p scattering, scattering by electrons, impurities and defects are also known to limit the phonon transport in metals. It is believed that the p-p scattering dominates over the others at $T > \Theta_{D}$ in general. In low electron density semiconductors and insulators, the p-e scattering rates are unimportant around or above $\Theta_{D}$ \cite{holland}. Nevertheless in heavily doped semiconductors and metals, the high electron density leads to strong electron-phonon coupling strength. The p-e scattering rates of these systems can be comparable to that of anharmonic p-p scatterings and will have strong impact on the thermal conductivity \cite{liao, jain}. In 2D materials like graphene, one can highly tune the carrier density and hence the electron-phonon coupling strength by external electric field. It is found that the e-ph coupling strength in graphene increases with the Fermi energy \cite{eph3}. For lower $E_F$, the coupling strengths are almost same both in electron and hole doping cases. Electron doping is predicted to have larger coupling strength than that of hole doping at higher $E_F$ \cite{eph3}.             

\begin{figure}
\centering
\includegraphics[scale = 0.32]{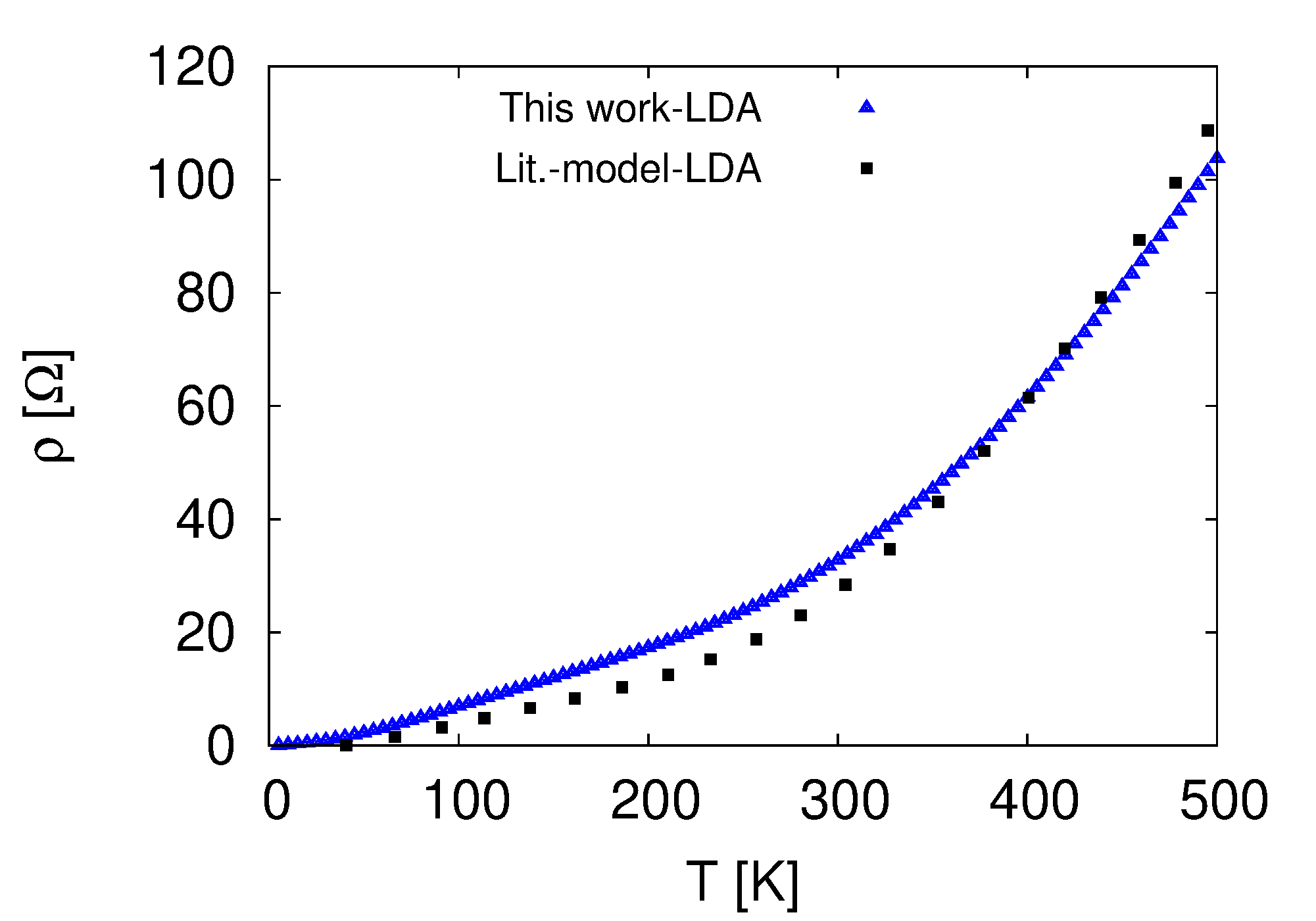}
\caption{Phonon limited intrinsic electrical resistivity of charge doped graphene ($n = 2.86 \times 10^{13}$ cm$^{-2}$). The literature results are taken from \cite{gra-res}}
\label{fig:res-gra}
\end{figure}

In this work we have studied the effect of electron doping on the lattice thermal conductivity of graphene by shifting the $E_F$ towards the unoccupied energy states from the charge neutral point. With external field one can achieve the carrier density in graphene as high as $n = 4 \times 10^{14}$ cm$^{-2}$ \cite{gra-expt}. To obtain such a large carrier density the corresponding shifting of $E_F$ will be $\sim$ 2.7 eV considering the Dirac point of neutral graphene at 0.0 eV. We have included many cases between the Dirac point and 2.7 eV in order to present a detailed overview. For every case, the scattering rates of phonons by electrons are calculated using Allen's model \cite{allen}. We would like to mention that we have compared our calculated electrical resistivity with the earlier report for a particular electron density ($n = 2.86 \times 10^{13}$ cm$^{-2}$). Our result matches well with the plot presented in reference \cite{gra-res} (see Fig.~\ref{fig:res-gra}). We believe this is beyond the scope of this manuscript. So, we are not going into the detailed discussion on the electrical resistivity.

\begin{figure}
\centering
\includegraphics[scale = 0.6]{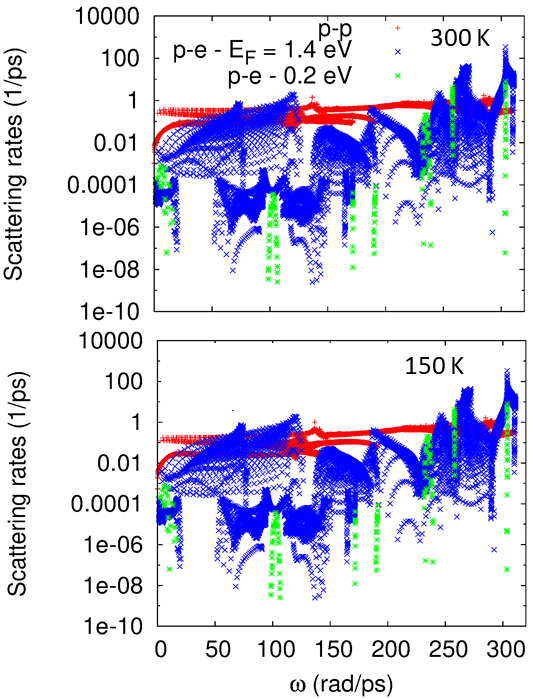}
\caption{Phonon-phonon anharmonic scattering rates of pristine graphene compared with phonon-electron scattering rates of electron doped graphene.}
\label{fig:gra-pp-pe}
\end{figure}

\begin{figure}
\centering
\includegraphics[scale = 0.6]{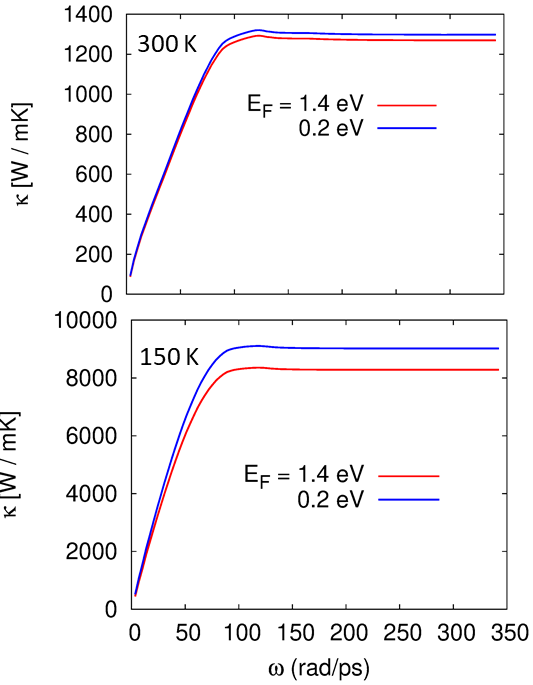}
\caption{Cumulative thermal conductivity of electron doped graphene compared at two different temperatures.}
\label{fig:gra-cum-kappa}
\end{figure}

The p-e scattering rates of pure and two electron doped cases are compared with the p-p anharmonic scattering rates in Fig.~\ref{fig:gra-pp-pe}. For low level doping, the p-e scattering rates are negligible while comparing with the p-p anharmonic one at low frequency. This signifies that the e-ph coupling strength is minimal for lower $E_F$ as reported in \cite{eph3}. The discrete scattering rates for $E_F$ = 0.2 eV are the outcome of intervalley scattering at K-point (see Fig.~\ref{fig:phdis}). Both acoustic and optical K-phonons are weakly scattered by intervalley scattering of electrons and holes at the Dirac point in graphene. With higher doping, the scattering becomes much stronger. Above 250 $rad/ps$ the lifetime of phonons for all level of doping significantly starts decreasing. But, as we will see in Fig.~\ref{fig:gra-cum-kappa} higher energy phonons contribute less in the thermal transport and hence stand as unimportant. Fig.~\ref{fig:gra-pp-pe} further suggests that the overlapping between the p-p and p-e scattering rates increases with lowering temperature ensuring strong effect on the heat transport. The same is reflected in the cumulative thermal conductivity which are plotted in Fig.~\ref{fig:gra-cum-kappa} for the same temperatures. As it can be seen from Fig.~\ref{fig:gra-cum-kappa} the effect of doping seems to have more impact at low temperature and decreases with rise in temperature. Only phonons that lie below 125 $rad/ps$ have contributions in the thermal transport. 

\begin{figure}
\centering
\includegraphics[scale = 0.32]{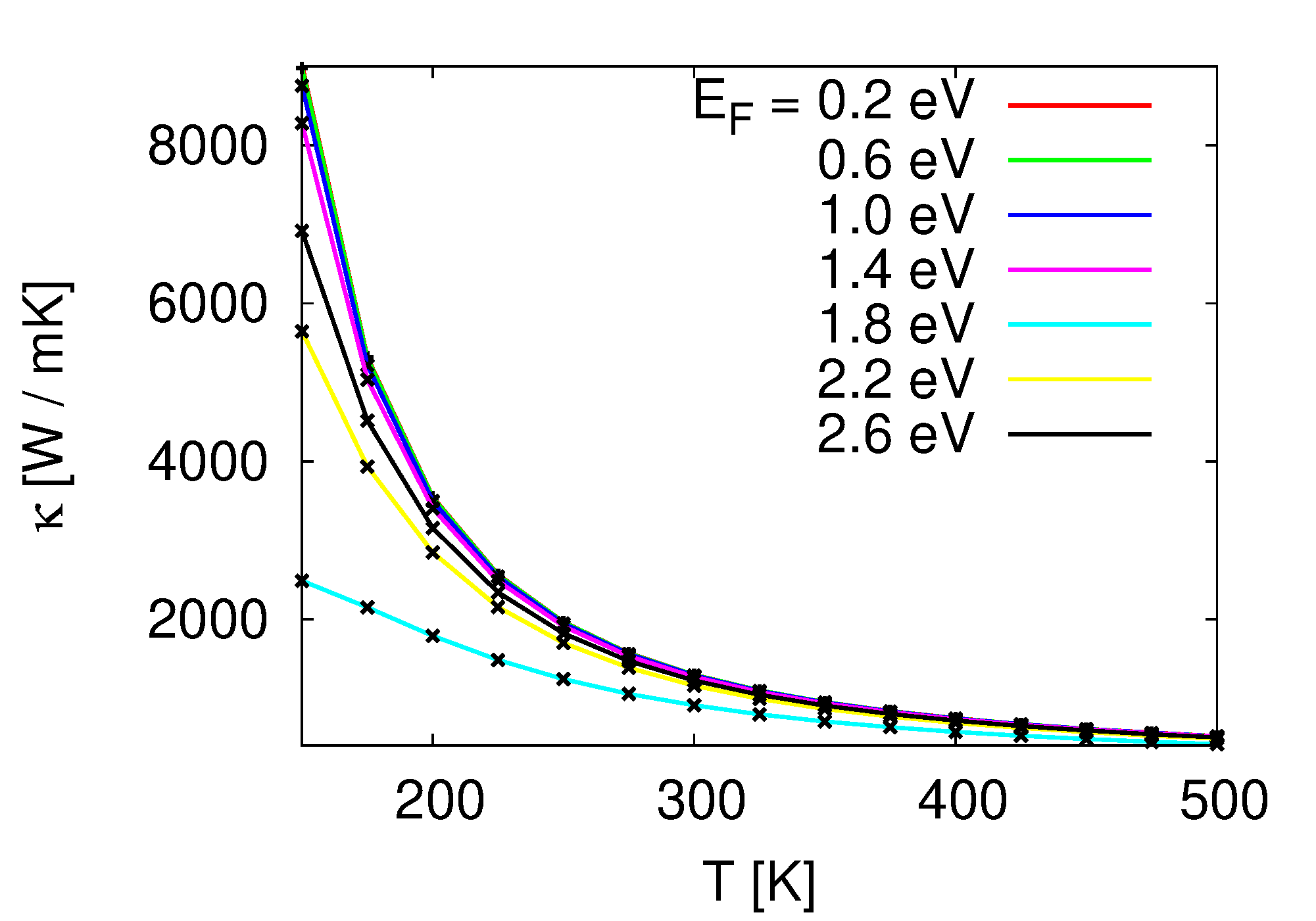}
\caption{Lattice thermal conductivity of doped graphene varies with temperature.}
\label{fig:gra-kapa-conv}
\end{figure}

\begin{figure}
\centering
\includegraphics[scale = 0.5]{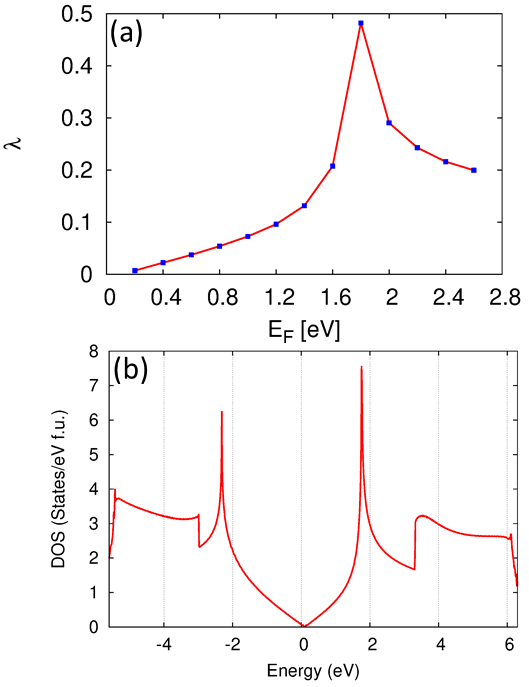}
\caption{(a) The electron-phonon coupling strength vs $E_F$ in graphene. (b) Density of states of pure graphene.}
\label{fig:lambda-dos}
\end{figure}

We now present the lattice thermal conductivity of doped graphene for a range of temperature. The effect of doping is seen to have decreasing trend with temperature. Moreover, as far as the doping level is concerned the predicted result is quite interesting and we believe it is an important finding in graphene, an extensive studied system. With an interval of 0.2 eV we have calculated the p-e scattering rates from $E_F$ = 0.2 to 2.6 eV and for some cases the corresponding lattice thermal conductivity are calculated which are shown in Fig.~\ref{fig:gra-kapa-conv}. The thermal conductivity does not vary monotonically w.r.t. the change in $E_F$. It starts decreasing with increase of $E_F$ till $E_F$ = 1.8 eV and then reverses it trend. Comparing the two cases, $E_F$ = 0.2 eV and 1.8 eV, the reduction of lattice thermal conductivity is found to be $\sim 50\%$ at 200 K. To understand the nature of this trend we have compared the calculated electron-phonon coupling strengths in doped graphene. The plot in Fig.~\ref{fig:lambda-dos}(a) explains satisfactorily the behaviour of lattice thermal conductivity w.r.t. the $E_F$. As we can see from Fig.~\ref{fig:lambda-dos} $E_F$ = 1.8 eV provides maximum coupling strength. Stronger the e-ph coupling strength stronger will be its effect on the lattice heat conduction. It is already shown in Fig.~\ref{fig:gra-pp-pe} that the p-e scattering rates at $E_F$ = 1.4 eV are larger than that at $E_F$ = 0.2 eV. Larger scattering rate is the signature of strong e-ph coupling which is maximum at $E_F$ = 1.8 eV. The coupling strength is directly related to the density of states (DOS) that is plotted in Fig.~\ref{fig:lambda-dos}(b). The DOS reaches its maximum value at $E_F$ = 1.8 eV and then starts decreasing.  


\section{\label{sec:level4}summary and conclusions}
In summary, we carry out \textit{ab initio} calculations to investigate the effect of charge doping on the lattice thermal transport of suspended graphene. The charge doping is incorporated by shifting the Fermi energy from the charge neutral Dirac point. Externally induced phonon-electron scattering rates are comparable to that of phonon-phonon scattering rates at large doping. We find that $\sim 50\%$ reduction of the lattice thermal conductivity can be achieved at 200 K. Our results open up new possibilities for potential thermoelectric applications of graphene in practical and for other 2D systems in general.

\section*{Acknowledgements}
A. Jena acknowledges the financial support from Shenzhen Science, Technology and Innovation Commission.

\bibliography{paper} 
\providecommand{\noopsort}[1]{}\providecommand{\singleletter}[1]{#1}%
\end{document}